\documentclass{iau_FM}
\usepackage{graphicx}
\usepackage{url}

\usepackage{natbib}

\title[SEP environment in the inner heliosphere] %% give here short title %%
{SEP environment in the inner heliosphere from Solar Orbiter and Parker Solar Probe}

\author[Wimmer-Schweingruber et al.]   %% give here short author list %%
{Robert F.\,Wimmer-Schweingruber$^1$
  \and Javier Rodriguez-Pacheco$^2$
  \and George C.\,Ho$^3$
  \and Christina M.\,Cohen$^4$
  \and Glenn M.\,Mason$^5$
  \and \\the Solar Orbiter EPD and Parker Solar Probe ISIS teams
}

\affiliation{$^1$Institute of Experimental \& Applied Physics\\Kiel University,\\24118 Kiel, Germany \\ email: {\tt wimmer@physik.uni-kiel.de} \\[\affilskip]
$^2$Universidad de Alcalá, Space Research Group, 28805 Alcalá de
  Henares, Spain\\email: {\tt javier.pacheco@uah.es } \\[\affilskip]
  $^3$Southwest Research Institute, San Antonio, TX, USA\\email: {\tt george.ho@swri.org}\\[\affilskip]
  $^4$California Institute of Technology, Pasadena, CA, USA\\email: {\tt cohen@srl.caltech.edu}\\[\affilskip]
  $^5$Applied Physics Laboratory, Johns Hopkins University, Laurel, MD 20723 USA\\email: {\tt glenn.mason@jhuapl.edu}
}

\pubyear{2024}
\jname{Astronomy in Focus, Volume 1} 
\editors{Piero Benvenuti, ed.}
\begin{document}

\maketitle

\begin{abstract}
The Sun drives a supersonic wind which inflates a giant plasma bubble in our very local interstellar neighborhood, the heliosphere. It is bathed in an extremely variable background of energetic ions and electrons which  originate from a number of sources. Solar energetic particles (SEPs) are accelerated in the vicinity of the Sun, whereas shocks driven by solar disturbances are observed to accelerate energetic storm particles (ESPs). Moreover, a dilute population with a distinct composition forms the anomalous cosmic rays (ACRs) which are of a mixed interstellar-heliospheric origin. Particles are also accelerated at planetary bow shocks. We will present recent observations of energetic particles by Solar Orbiter and Parker Solar Probe, as well as other spacecraft that allow us to study the acceleration and transport of energetic particles at multiple locations in the inner heliosphere.
%, {\bf I don't have the file Keywords.txt!!!}}
\end{abstract}

\firstsection % if your document starts with a section,
\section{Introduction}
\label{sec:intro}

The Integrated Science Investigation of the Sun \cite[IS$\odot$IS, ][]{mccomas-etal-2016} on Parker Solar Probe \cite[PSP, ][]{fox-etal-2016} and the Energetic Particle Detector \cite[EPD, ][]{rodriguez-pacheco-etal-2020} on Solar Orbiter \cite[]{mueller-etal-2020} basically have the same science goals which are worded slightly differently and are summarized here for convenience. \cite{mccomas-etal-2016} states IS$\odot$IS' top level science goal as to determine the physical mechanisms that produce, accelerate, and transport energetic particles in the inner heliosphere. As usual, this is broken down into three lower-level science questions: a) What is the origin or seed population of solar energetic particles (SEPs)? b) How are these SEPs and other particle populations accelerated? c) What mechanisms are responsible for transporting the different particle populations into the heliosphere? On the other hand, EPD 's top-level science question is \cite[see ][]{rodriguez-pacheco-etal-2020} to determine how solar eruptions produce energetic particle radiation that fills the heliosphere. This is also broken down into three more detailed science questions. a) How and where are energetic particles injected at the sources and, in particular, what are the seed populations for energetic particles? b) How and where are energetic particles accelerated at the Sun and in the interplanetary medium? c) How are energetic particles released from their sources and distributed in space and time? 

Obviously there are three common science questions which can be broken down into three processes which are needed for energetic particles to be measured in the heliosphere: injection (sec.~\ref{sec:injection}), acceleration (sec.~\ref{sec:acceleration}), and transport (sec.~\ref{sec:transport}). We have chosen to structure this paper according to these three processes and to provide examples for these processes that use data from Solar Orbiter, PSP, and other spacecraft that were recently reported. Because we have been tasked by the organizers of this workshop with providing an overview of results from PSP and Solar Orbiter, this paper does not report brand new results. Rather, it attempts to exemplify how the new instruments on these missions as well as the various constellations offered by them and other inner-heliosphere missions can elucidate these three important contributions to accelerating energetic particles that fill the heliosphere. As usual, we will provide a summary and discussion at the end of the paper.

\section{Injection}
\label{sec:injection}

In order to do their mischief in the heliosphere, energetic particles need to be injected into the acceleration process. If they are accelerated at a shock, their speed needs to exceed the thermal speed of the local particle population so that they can freely cross the acceleration region (e.g., a shock) multiple times. Just how they attain this elevated kinetic energy has been a highly debated topic in the past decades. Today we know that the solar wind does not serve as the sole or main reservoir for suprathermal particles but that there appears to be a population of such particles available for further acceleration. An important clue came from measurements of the composition of suprathermal particles. They are enriched in the rare $^3$He isotope throughout the solar activity cycle (but do show a dependence on it, \cite{wiedenbeck-and-mason-2013}). \cite{desai-and-giacalone-2016} provide an excellent overview of the measurements and their consequences for interpretation.

\begin{figure}
  \centering
  \includegraphics[width=0.98\textwidth]{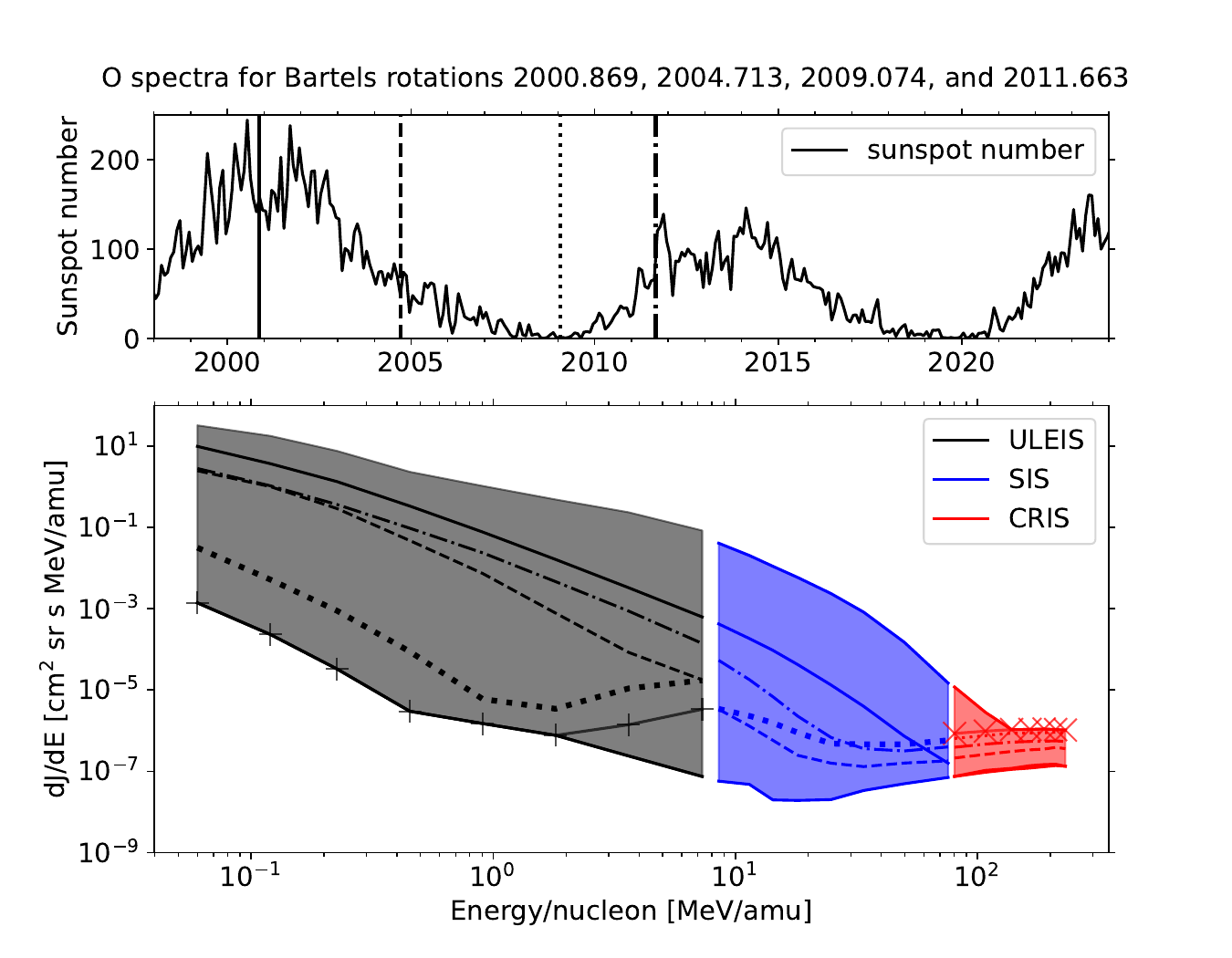}
  \caption{Upper panel: Sunspot numbers vs.\,time (from WDC-SILSO, Royal Observatory of Belgium, Brussels).  Bottom panel: Oxygen differential flux spectra accumulated over the time periods indicated by vertical lines in the upper panel (decimal years 2000.869, 2004.787, 2009.074, and 2011.663). Line styles in the lower panel indicate the Bartels rotation of the observation as indicated in the upper panel. Colors indicate different instruments on the ACE spacecraft as discussed in the text. The shaded region shows the envelope of the variability of the Bartels-rotation-averaged O particle fluences from 1998 -- 2024. See text for an explanation of the ``x'' and ``+'' symbols. }
  \label{fig:variable-1}
\end{figure}

As is well known, the Sun is an active star with an activity cycle driven by the solar dynamo \cite[]{cliver-etal-2022}. This is illustrated in the upper panel of Fig.~\ref{fig:variable-1} which shows sunspot number (Source: WDC-SILSO, Royal Observatory of Belgium, Brussels) which is a good indicator for solar activity as a function of time.
The bottom panel shows oxygen differential flux spectra accumulated over the four Bartels rotations indicated by vertical lines in the upper panel and in the title of the Figure. The line styles in the lower panel correspond to those in the upper panel. Particle data were taken from three instruments on NASA's Advanced Composition Explorer (ACE, \cite{stone-etal-1998}), the Ultra-Low-Energy Isotope Spectrometer (ULEIS, \cite{mason-etal-1998}) the Solar Isotope Spectrometer (SIS, \cite{stone-etal-1998b}), and the Cosmic-Ray Isotope Spectrometer (CRIS, \cite{stone-etal-1998c}).
%ACE/CRIS data for Bartels rotation 2001.830 (dotted line) are discussed in the following paragraph.

Fig.~\ref{fig:variable-1} also shows the highest and lowest particle fluences observed during the course of the ACE mission and illustrates the enormous variability of particle fluxes which is bounded by the shaded area. It covers a dynamic range of about seven orders of magnitude even when averaged over entire Bartels rotations. Some further explanations are needed to interpret the figure. Fluxes at the  highest energies covered by ULEIS during quiet times are affected by instrumental noise which is due to a malfunction of the second time-of-flight (ToF) measurement. The measured values (from the ACE Science Center) are plotted as a line plot with a ``+'' symbol. Because we know that the lower limits are not correct over the entire energy range, we have also plotted a logarithmic interpolation between the lowest-energy SIS data point and the ULEIS data point with the highest energy which is not affected by this background (around 1 MeV/nuc). Moreover, the CRIS instrument on ACE was not designed to measure high fluxes during solar particle events but to provide accurate measurements of the quiet-time galactic cosmic rays. It saturates during high-flux time periods\footnote{Dr.\,R.\,Leske, Caltech, private communication, 2022. See also the caveat at \url{https://izw1.caltech.edu/ACE/ASC/level2/cris_l2desc.html} which states that ``Time periods during which solar activity is high are not covered by CRIS, since the instrument was not designed to operate during such conditions.''} which is seen by the measured data points which are also shown as a line plot with an ``x'' symbol marking. To establish a likely envelope, shown as the solid line bounding the shaded region, we have used a fluence of 0.8 times the highest-energy SIS data point and 0.8 times the square root of that value multiplied by the second-lowest CRIS fluence value. This is also the reason for the ``missing'' CRIS data in Fig.~\ref{fig:variable-1}. Figure~\ref{fig:variable-1} and its discussion above are updated versions of a similar figure shown in \cite{wimmer-schweingruber-etal-2023}.

Here we illustrate the injection stage with new measurments performed with the Suprathermal Ion Spectrograph \cite[SIS][]{rodriguez-pacheco-etal-2020} on Solar Orbiter. SIS measures total energy of and performs multiple time-of-flight measurements on suprathermal particles. This results in a very low background which allows measurements of the heliospheric quiet-time background of suprathermal particles \cite[]{mason-etal-2021, mason-etal-2023}. We will not consider those measurements here but rather investigate the origin of particles injected into and accelerated during two impulsive solar particle events observed by Solar Orbiter on the 9th of November 2022 and on the 8th of April 2023, as reported by \cite{mason-etal-2023}. The first of these events was very unusual in that it contained several hundreds of ultraheavy ions (with masses $>78$ amu). The second event occurred during very low ambient activity which allowed those authors to observe particle spectra with only minimal pre-event background from the suprathermal seed population. These observations showed very low proton and $^4$He intensities together with intensities of other ions that are typical of $^3$He-rich events. Together these observations indicate that the enrichments in Fe/O, $^3$He/$^4$He, etc.\,which have often been attributed to the injection process are more likely due to the acceleration process itself. In the model of \cite{roth-and-temerin-1997} and \cite{temerin-and-roth-1992} ions are accelerated by electromagnetic hydrogen cyclotron waves in a single-stage process. Ions are heated over a broad range of gyrofrequencies, but in the model presented by \cite{mason-etal-2023} the spectrum of waves is strongly damped at the gyrofrequencies of the abundant H and $^4$He. Thus the ``unusual'' abundance features such as enhanced Fe/O is not due to an enrichment of Fe but rather to a depletion of O because it is fully stripped and thus in resonance with the waves damped by $^4$He. This process would also explain the enrichments of the neutron-rich isotopes of Ne, Mg, Si, etc.\,which have been reported in the past \cite[]{dwyer-etal-2001, leske-etal-2001, leske-etal-2007, mason-etal-1994}.

\section{Acceleration}
\label{sec:acceleration}

Acceleration of particles requires work to be done on them which in turn requires an electric field. Strong electric fields are difficult to maintain in a highly conductive plasma such as the corona or solar wind. Maxwell's equations allow a way out of this dilemma by inducing electric fields with a rapidly changing magnetic field. Rapid temporal variations can be achieved in solar flares as a consequence of reconnection. Rapid spatial variations such as shocks or other magnetic discontinuities translate to rapid temporal changes as they are convected into a plasma. Finally waves traveling through the plasma can also heat particles which are in resonance. \cite{yang-etal-2023} and \cite{trotta-etal-2023} have investigated the acceleration of suprathermal ions using data from the SupraThermal Electron Proton (STEP) sensor which is part of EPD on Solar Orbiter and offers unusually high energy and time resolution \cite[]{rodriguez-pacheco-etal-2020, wimmer-etal-2021, yang-etal-2023}. 

\cite{yang-etal-2023} reconstructed the pitch-angle distributions (PADs) of suprathermal ions in the solar wind frame of reference and studied its temporal evolution, as well as the flux profile and velocity distribution function of these particles in the vicinity of a shock observed on the 3rd of November 2021. They found that suprathermal proton fluxes peaked $\sim 12$ to $\sim 24$s before the shock in the upstream region, and that close to the shock they exhibited anisotropic PADs that pointed away from the shock and tended to isotropize farther upstream. Downstream of the shock, the PADs showed anisotropies towards a pitch angle of 90$^\circ$. Comparison of their measurements with theoretical models suggests that the acceleration of suprathermal protons at interplanetary shocks is highly dynamic on a timescale of $\sim 10$s, i.e., few proton gyroperiods. Furthermore, they conclude that shock-drift acceleration likely plays an important role in accelerating these suprathermal protons. They suggest that the strong variability of the suprathermal particle fluxes could be a result of time-varying acceleration efficiency at the shock or that it is a consequence of shock rippling \cite[]{johlander-etal-2016, johlander-etal-2018, yang-etal-2018}, i.e., a rapidly varying magnetic connection to different parts of the shock that would have different local $\theta_{Bn}$ and therefore different acceleration efficiencies.

\cite{trotta-etal-2023} combined results from hybrid kinetic simulations with STEP measurements of suprathermal ions in the vicinity of the 30 October 2021 shock to show that the inherent variability of the injection process in both time and space must be considered to solve the problem of how suprathermal particle injection occurs in astrophysical systems. In other words, their results strongly support the second explanation given by \cite{yang-etal-2023}. Similar to that work, they found highly variable particle enhancements but in this case they showed clear velocity dispersion in the vicinity of the shock on time scales of 10 to 20 s. In their view, the particles are accelerated at different locations of the shock which are magnetically connected to the spacecraft at different times. These authors also used 2.5-dimensional kinetic simulations to model details of the shock and particle heating and acceleration. They found that suprathermal particles are not distributed uniformly in the vicinity of the shock. Their flux varies strongly along the shock front which indicates irregular injection into the acceleration process.

Together the results of \cite{yang-etal-2023} and \cite{trotta-etal-2023} underline the importance of detailed, high-resolution studies in the vicinity of interplanetary shocks to understand both the injection and acceleration of particles. They also drive home the point that injection (Sec.~\ref{sec:injection}) and acceleration (Sec.~\ref{sec:acceleration}) are not isolated processes but need to be considered together. We could just as well have presented these results in Sec.~\ref{sec:injection}!

\section{Transport}
\label{sec:transport}

Particles can escape the injection/acceleration region after which we speak of ``transport''. A simple thought experiment starting with a point-like source may serve to illustrate this process. If we assume a point-like source then particles can propagate along the interplanetary magnetic field (IMF), they can experience drifts and diffusion perpendicular to the IMF, and further energy and momentum changes due to changes in the IMF at all temporal and spatial scales. Finally, flux tubes filled with energetic particles to different levels can be convected by an observer which results in interesting temporal behavior \cite[see, e.g., ][]{wimmer-etal-2023}. To illustrate these effects we will discuss four publications by \cite{trotta-etal-2024}, \cite{telloni-etal-2021c}, \cite{kollhoff-etal-2021}, and \cite{mason-etal-2021}.

How can we decide whether an observed feature in a particle event is due to injection/acceleration or due to transport? To answer that we need to know how the source, e.g., a shock evolves in the inner heliosphere where it accelerates particles. \cite{trotta-etal-2024} observed a CME-driven interplanetary shock at 0.07 with Parker Solar Probe and at 0.7 with Solar Orbiter on the 5th of September 2022. This allowed them to study the shock properties at these two different heliocentric distances. Coincidentally, the 0.07 au or 15 $R_\odot$ are also the closest to the Sun that a shock has been observed to date. PSP (Solar Orbiter) was located at Carrington longitude 232$^\circ$ (252$^\circ$) and latitude -1.8$^\circ$ (-3.6$^\circ$). While this is not a perfect alignment the measured shock properties were nevertheless very similar at the two locations: $\theta_{Bn}$ was 53$^\circ$ (51$^\circ$), shock magnetic compression ratio, $r_B$, was 2.3 (1.9), and the gas compression ratios was 1.6 (2.0). Shock speed decreased from 1520 km/s at PSP to 942 km/s at Solar Orbiter, but the fast magnetosonic and Alfvénic Mach numbers were very similar ($M_{fms} = 3.8\ \ (3.2)$ and $M_A = 3.9 \ \ (3.8)$). Despite these similarities, the shock at PSP was seen as a sharp transition without any signs of an upstream foreshock, but much more complex at Solar Orbiter. There a wavelet spectrum of the magnetic ﬁeld fluctuations shows enhanced power extending to small ($\sim$1 Hz) scales indicating upstream shocklet activity. At PSP the downstream level of fluctuations is enhanced by a factor $\sim$4 with respect to the upstream levels, whereas at Solar Orbiter this region is populated with strong compressive and non-compressive magnetic ﬁeld ﬂuctuations, which indicates that the shock propagated through a much more structured portion of the solar wind than at PSP. These authors conclude that the shock environment at Solar Orbiter is much more disturbed than at PSP.

\cite{telloni-etal-2021c} used the ﬁrst radial alignment between Parker Solar Probe and Solar Orbiter to investigate the
evolution of solar wind turbulence in one and the same plasma parcel in the inner heliosphere. They allowed a tolerance of $\pm$1.5 degrees in latitude and longitude to identify a 1.5 hour time period centered around 04:00 UT on the 27th of September 2020 at PSP (00:00 UT on the 2nd of October 2020 at Solar Orbiter) as corresponding to the same plasma parcel at 0.1 and 1 au. Using burst-mode magnetic field measurements at the two spacecraft, they could show that the two spacecraft indeed measured very similar characteristics of the solar wind stream. The stream originated from a quiescent region and showed exceptionally smooth magnetic field and a Parker-like field orientation without any switchbacks. They used standard diagnostics for spectral power, compressibility, and intermittency of the magnetic ﬁeld ﬂuctuations to show that at PSP turbulence was still highly Alfénic while it showed fully developed, intermittent turbulence at Solar Orbiter. They concluded that nonlinear interactions between Alfvén waves probably did not yet have time to fully develop at 0.1 au, where the solar wind sampled by PSP was still pristine and unaffected by stream–stream interactions. At Solar Orbiter (at 1 au) the same plasma has  evolved considerably, the turbulent cascade has fully developed and Alfvénic ﬂuctuations have decreased. 

\cite{kollhoff-etal-2021} investigated the solar particle event observed by multiple spacecraft (ACE, Parker Solar Probe, SOHO, Solar Orbiter, and STEREO-A) across a longitudinal range of $> 230^\circ$ on the 29th of November 2020. This was the first wide-spread event of solar cycle 25 and accelerated electrons to near-relativistic energies as well as ions to energies $> 50$ Mev. It was associated with an M4.4 class X-ray flare, a coronal mass ejection and an extreme ultraviolet (EUV) wave as well as a type II radio burst and multiple type III radio bursts. \cite{kollhoff-etal-2021} used velocity dispersion and time-shift analysis techniques to identify the particle release times at the Sun to show that the delays in onset time increased with increasing angle between the flare foot point and the observer. They compared the thus determined onset times with the time the EUV wave intersected the estimated magnetic footpoints of each spacecraft and concluded that a simple scenario where the particle release is only determined or triggered by the EUV wave is unlikely. Anisotropic particle distributions were observed at SolO, Wind, and STEREO-A. Remarkably, they found that the largest anisotropy was observed by STEREO-A which was separated by $\sim 95^\circ$ from the parent solar active region. Their observations suggest that particles are injected over a wide longitudinal range close to the Sun, in fact even crossing the heliospheric current sheet. 

\cite{mason-etal-2021} investigated the same event (29th of November 2021) as \cite{kollhoff-etal-2021} in view of the heavy ion properties of energetic particles. SIS and ULEIS on ACE (which is a very similar instrument to SIS) saw only upper limits of $< 1\%$ and $0.03\%$ to the enrichment of $^3$He in this event. This could simply be a consequence of this being the first large solar particle event of solar cycle 25 and that the ``reservoir'' of suprathermal flare material \cite[]{wiedenbeck-and-mason-2013} had not yet been replenished enough to provide sufficient $^3$He seed particles. Heavy ion abundances normalized to oxygen were observed to be about two time lower at PSP and ACE than long-term averages \cite[]{desai-etal-2006}. PSP observed a nearly monotonic drop in Fe/O with increasing energy, consistent with measurements at STEREO and ACE. At low energies (below $\sim 30$ keV), however, PSP measured a significantly enhanced Fe/O ratio that increased with decreasing energy. These measurements extend below the energy range of most previous SEP events and may be a new feature.

\section{Summary and Discussion}
\label{sec:summary}

We have considered several new results from Solar Orbiter and Parker Solar Probe that illustrate the complexity of the question being tackled by the energetic particle instruments on these two spacecraft. The question ``How do solar eruptions produce energetic particle radiation that fills the heliosphere?'' sounds deceitfully simple but reveals a remarkable richness in phenomena once one digs into the details. We have broken it down into three stages which can bring an energetic particle to end up in one of the many instruments used to measure such radiation: injection, acceleration, and transport. We have seen that these different stages are strongly interlinked and are not easily separated. The often reported enhancements of Fe/O were shown by \cite{mason-etal-2023} to be due to a depletion of fully stripped O, at least for the two events discussed in that publication. They propose that the waves responsible for heating particles are strongly damped by protons and $^4$He and thus there is less wave power available to ions with charge-to-mass ratios close to 0.5. This interpretation implies that not all variability in heavy-ion abundances (generally normalized to O) would be due to enrichments of those elements in the ``seed population'', but that the acceleration mechanism itself causes these apparent enrichments. Therefore the distinction between ``injection'' and ``acceleration'' becomes blurred and in a sense ill-defined. 

The STEP sensor, part of Solar Orbiter's EPD suite of energetic particle instruments, is capable of measuring at an exceptionally high cadence of one second. \cite{yang-etal-2023} and \cite{trotta-etal-2023} used this 15-pixel pin-hole instrument to reconstruct high-resolution pitch-angle distributions in the vicinity of two shocks. Their high time-resolution results allowed them to catch shock-drift acceleration in the act \cite[]{yang-etal-2023} and revealed bursty, non-local acceleration close to IP shock \cite[]{trotta-etal-2023}. STEP measures in a previously hardly accessible energy range close to the bulk energy of the solar wind. The remarkable dynamics in the vicinity of the shocks again demonstrate how blurred the distinction between  ``injection'' and ``acceleration'' has become. 

Once injected and accelerated, energetic particles leave their source region and experience ``transport'' effects. We may have been a little too inclusive here by also including the radial evolution of shocks in Sec.~\ref{sec:transport} (Transport) as discussed by \cite{trotta-etal-2024}, but this is an important effect. Shocks do evolve with distance from the Sun and the new constellations of spacecraft in the inner heliosphere allow us to measure this evolution. Similarly, \cite{telloni-etal-2021c} measured the radial evolution of turbulent fluctuations in a plasma parcel between 0.1 and 1 au. This went beyond the usual statistical sampling by identifying one and the same plasma parcel at PSP and Solar Orbiter. \cite{kollhoff-etal-2021} and \cite{mason-etal-2021} investigated the same event (29th of November 2021) at multiple spacecraft to elucidate especially the longitudinal spread of particles and the properties of heavy ions which serve as tracers for micro-physical processes in the injection/acceleration regions. 

The constantly changing constellations of inner-heliosphere spacecraft (including spacecraft at one au) are allowing us to better understand radial and longitudinal gradients and spreads of particles and properties of the solar wind and interplanetary magnetic field through which they propagate. Solar Orbiter leaving the ecliptic in 2025 will add latitudinal transport the equation. 

\bibliographystyle{habil}
\bibliography{no-duplicates-diss.bib} 

\begin{discussion}

\discuss{Bernd Heber}{ How certain can it be that we see the same plasma parcel with PSP and SOLO (Telloni paper)?}

\discuss{Robert Wimmer-Schweingruber}{It is certainly not exactly the same plasma parcel, but we took care to identify the same solar wind stream. That may differ across a few degrees in longitiude, but note that the changes we reported are many orders of magnitude, So the result is not affected by those small local variations.}

\discuss{Vladimir Airapetian}{ Can you also measure rarer elements such as potassium? This would be important for FIP studies.}

\discuss{Robert Wimmer-Schweingruber}{Yes, in principle SIS has the resolution to measure these rare ions but we have not yet performed such studies. }

\discuss{Nariaki Nitta}{How many circumsolar events have we seen and what have we learned from them?}
\discuss{Robert Wimmer-Schweingruber}{With the now available constellation of inner-heliosphere spacecraft we have seen several such events, but sofar studies have focussed on individual events. I am not aware of a study that compares such events or even presents statistical properties of them.}

\end{discussion}

\end{document}